\documentclass[]{spie}  
\usepackage{appendix}
\usepackage[]{graphicx}
\usepackage{float}
\usepackage{multirow}
\usepackage{mathtools}
\graphicspath{ {./figures/} }

\newcommand{\tabcell}[1]{\begin{tabular}{@{}c@{}}#1\end{tabular}}

\newcommand{\arcsec}{''}

\title{The InfraRed Imaging Spectrograph (IRIS) for TMT: photometric characterization of anisoplanatic PSFs and testing of PSF-Reconstruction via AIROPA} 

\author{
Nils Rundquist\supit{a,b},
Shelley A. Wright\supit{a,b}, 
Matthias Schoeck\supit{f,g}, 
Arun Surya\supit{a},
Jessica Lu\supit{c},
Paolo Turri\supit{i},
Edward L. Chapin\supit{f}, 
Eric Chisholm\supit{g}, 
Tuan Do\supit{d}, 
Jennifer Dunn\supit{f}, 
Andrea Ghez\supit{d},
Yutaka Hayano\supit{h} , 
Chris Johnson\supit{c}, 
James E. Larkin\supit{d}, 
Reed L. Riddle\supit{e}, 
Ji Man Sohn\supit{d}
Ryuji Suzuki\supit{h}
Gregory Walth\supit{j}
Andrea Zonca\supit{k}
\skiplinehalf
\supit{a} Center for Astrophysics \& Space Sciences, University of California San Diego, USA; \\
\supit{b} Department of Physics, University of California San Diego, USA; \\
\supit{c} Physics \& Astronomy Department, University of California Berkeley, Berkeley USA; \\
\supit{d} Physics \& Astronomy Department, University of California Los Angeles, CA 90095 USA; \\
\supit{e} Caltech Optical Observatories, 1200 E California Blvd., Pasadena, CA 91125 USA; \\
\supit{f} National Research Council of Canada - Herzberg, Victoria, BC, V9E 2E7 Canada; \\
\supit{g} Thirty Meter Telescope Observatory Corporation, Pasadena, CA 91105 USA; \\
\supit{h} National Astronomical Observatory of Japan, Osawa, Mitaka, Tokyo, 181-8588 Japan; \\
\supit{i} Department of Physics and Astronomy, University of British Columbia, Vancouver, BC V6T 1Z1 Canada \\
\supit{j} Carnegie Observatories, 813 Santa Barbara St, Pasadena, CA 91101 USA \\
\supit{k} San Diego Supercomputer Center, University of California San Diego, USA; 
}

\authorinfo{Further author information: (Send correspondence to N.R.)\\N.R...: E-mail: nrundqui@physics.ucsd.edu}

\begin{document} 
  \maketitle 

\begin{abstract}

The InfraRed Imaging Spectrograph (IRIS) is a first-light instrument for the Thirty Meter Telescope (TMT) that will be used to sample the corrected adaptive optics field by the Narrow-Field Infrared Adaptive Optics System (NFIRAOS) with a near-infrared (0.8 - 2.4 $\mu m$) imaging camera and integral field spectrograph.  To better understand IRIS science specifications we use the IRIS data simulator to characterize relative photometric precision and accuracy across the IRIS imaging camera 34\arcsec x34\arcsec field of view. Because the Point Spread Function (PSF) varies due to the effects of anisoplanatism, we use the Anisoplanatic and Instrumental Reconstruction of Off-axis PSFs for AO (AIROPA) software package to conduct photometric measurements on simulated frames using PSF-fitting as the PSF varies in single-source, binary, and crowded field use cases. We report photometric performance of the imaging camera as a function of the instrumental noise properties including dark current and read noise. Using the same methods, we conduct comparisons of photometric performance with reconstructed PSFs, in order to test the veracity of the current PSF-Reconstruction algorithms for IRIS/TMT.

\end{abstract}


\keywords{infrared:imaging, data:simulator, instrumentation: near-infrared, imaging:photometric, giant segmented mirror telescopes: Thirty Meter Telescope}


\section{Introduction}
\label{sec:intro}
\subsection{IRIS for TMT}

\begin{figure}[H]
    \centering
    \includegraphics[scale=0.52]{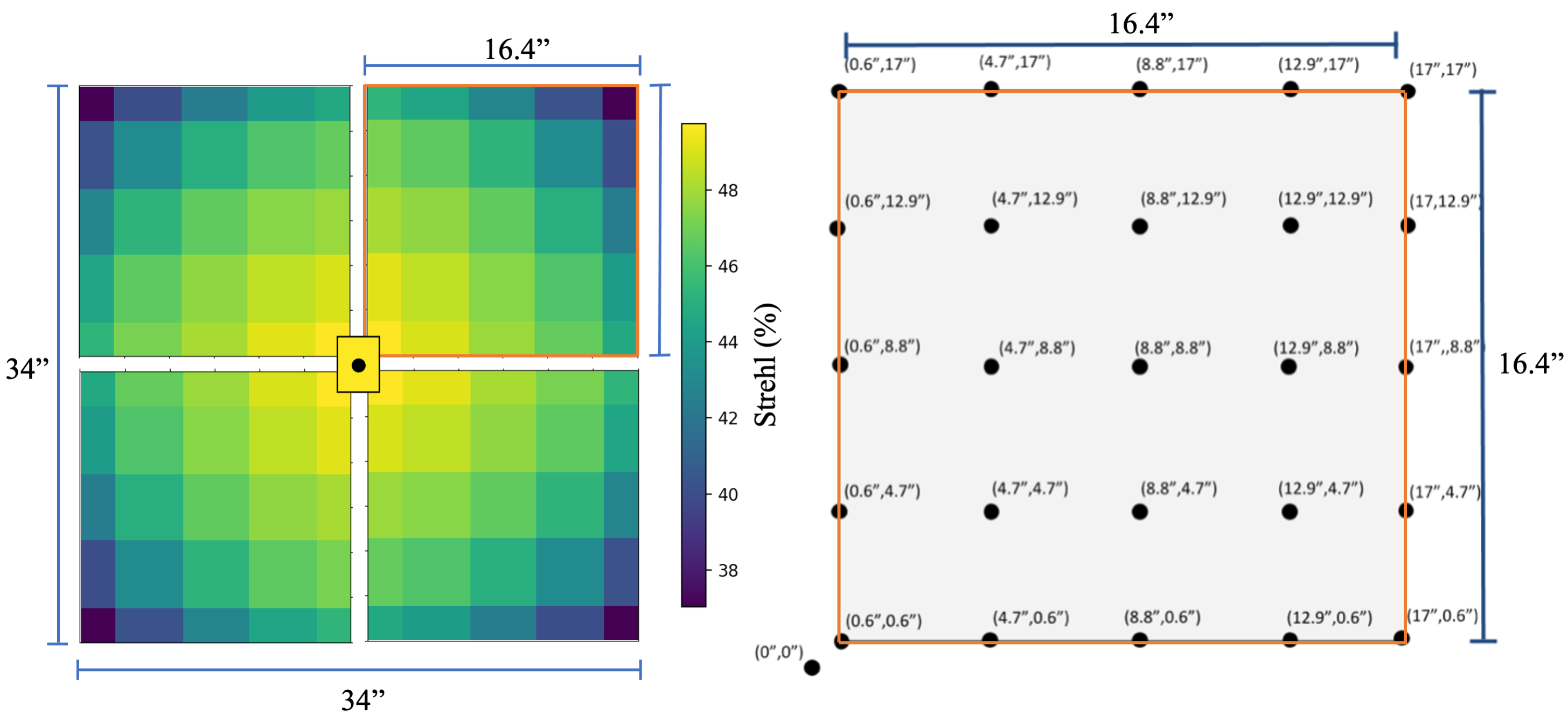}
    \caption{(Left) The setup of the IRIS imager detectors relative to the location of the spectrograph pick-off mirrors at the center of the four detectors, with the estimated Strehl ratio for K-band observations shown in color. The Strehl ratio for this figure was estimated assuming optimum performance of NFIRAOS and 75\% (best 75 percent, below average) atmospheric conditions. The IRIS simulations for photometric characterization were conducted over the upper-right imager detector relative to the on-axis position of the spectrograph pick-off mirror at the center of the IRIS focal plane, marked with the black dot for reference. (Right) The relative positions of the anisoplanatic PSFs used in the photometry simulations. When simulating a point source within the imager field of view, a unique PSF is generated for that field location by interpolating between the surrounding points from the above grid. The PSF grid itself is then used as the PSFs used for fitting by AIROPA, or interpolated to a finer-sampled grid for fitting as discussed in Section \ref{sec:crowded_field}
    \label{fig:iris_layout}}
\end{figure}

The InfraRed Imaging Spectrograph (IRIS)\cite{Larkin1,Larkin2,Larkin3,Larkin4} for the Thirty Meter Telescope (TMT) is an imaging camera (imager) and integral field spectrograph (IFS) operating in the near infrared 0.8 $\mu m$ - 2.4 $\mu m$ wavelength range.  The imager will sample the diffraction limit of TMT's 30 meter aperture and will provide a spatial plate scale of 4 miliarcseconds (mas) per spatial pixel over a 34 x 34 arcsecond field of view, distributed over four Teledyne Hawaii-4RG-10$\mu m$ detectors.  The IFS is comprised of both lenslet and slicer spectrographs that are capable of operating with varying spatial scales that operate in-parallel with the imager. The relative positions in the focal plane of the imager detectors and the spectrograph fields of view are illustrated in Figure \ref{fig:iris_layout}. IRIS will provide unprecedented diffraction-limited spatial resolution and characterizing its photometric performance is an important step in the development of IRIS science.

For the purposes of understanding the photometric capabilities of IRIS, we characterize the performance of the IRIS imager in terms of its relative precision and accuracy. The quality of the imager photometric performance is a function of the Signal-to-Noise Ratio (SNR) of an observed source within the field of view, as well as the observational parameters, atmospheric quality, and performance of the Narrow-Field Infrared Adaptive Optics System (NFIRAOS)\cite{adc1,adc2}.  To simulate these effects we use the IRIS data simulator\cite{Wright1,Wright2,Wright3} package in conjunction with imager Point Spread Functions (PSFs) simulated by the NFIRAOS team to indicate Strehl ratio, spatial atmospheric effects, and spatial asymmetries in flux distribution. Because the field of view of the imager is large relative to the spatial effects of atmospheric turbulence across the field of view, variations in the PSF caused by anisoplanatism are expected even with the best possible performance of NFIRAOS. Characterizing anisoplanatic effects is therefore an essential step in understanding the science capabilities of IRIS.

\subsection{PSFs}
\label{sec:psfs}
The PSF spatial variations across the imager field of view is dependent on wavelength and distance from field locations where the corrective effects of NFIRAOS are most pronounced. For our simulations we adopt a field configuration where the best AO performance is achieved at the on-axis location of the spectrograph pick-off mirror. For further descriptions of field locations relative to this position, we define the on-axis pick-off mirror location to be the origin of the imager-plane coordinate system (0", 0"), using arcseconds as our coordinate units. The data set of PSFs used in photometry simulations extends to the upper right imager detector corner location (17", 17") in a 5 x 5 grid starting at the lower left imager corner location (0.6", 0.6").  This grid of PSFs is shown in Figure \ref{fig:iris_layout}. For the sake of practicality in the scope of photometric simulations measuring the effects of anisoplanatism, we assume similar photometric performance in the other three imager detectors due to rotational symmetry given adequate AO performance relative to their positions. 

Using the PSFs described, we create a PSF unique to any field location by using a bilinear interpolation scheme between the closest four PSFs for a given location on the imager detector.  We then convolve the interpolated PSF with the flux for a given point source magnitude, and simulate the sky-background and instrumental effects of IRIS using the data simulator to approximate an observation. For photometry, we then use PSF-fitting routines to retrieve estimated flux values for each simulated source. 

\subsection{PSF Fitting Routines}
\label{sec:psf_fitting}
PSF-fitting routines are a necessary step in achieving high-precision measurements in AO data in order to account for extended spatial flux distribution as described by the PSF.  Aperture photometric methods are insufficient for achieving high precision measurements of crowded fields as shown by years of work performed with the \textit{DAOPHOT} algorithm and later by \textit{Starfinder}\cite{starfinder}. \textit{Starfinder/DAOPHOT} algorithms are effective tools in fitting an empirical or prescribed PSF to stellar fields whose PSF variability across the field is negligible, but this is not the case for the IRIS imager due to the high difference in Strehl ratio across the field of view shown in Figure \ref{fig:iris_layout} as a result of anisoplanatism. We then use the Anisoplanatic and Instrumental Reconstruction of Off-axis PSFs for AO (AIROPA)\cite{airopa, airopa2} software package in order to account for the anisoplanatic effect and subsequent variability in the PSF across the imager field of view. We use aperture photometry as a comparison for the results of the PSF-fitting routines provided by AIROPA. We also test the veracity of PSF-Reconstruction (PSF-R) algorithms, through using a separate data set of reconstructed PSFs for fitting with AIROPA; this is discussed in Section \ref{sec:PSFR}.

In conducting these simulations we not only hope to characterize the IRIS imager performance, but to use practical methods astronomers would use on the acquired data to extract source positions and photometric counts using PSF-fitting astrometry and photometry. Large detector fields of view such as is provided by the Hawaii-4RG detectors of the IRIS imager provide new problems for data processing beyond anisoplanatic PSFs.  Because the PSF varies across the field as described in \ref{sec:psfs}, we use AIROPA to fit PSFs in a field-variable manner. However, due to the large field of view and potentially large number of sources to be processed by AIROPA, CPU draw and processing time becomes a potential issue with very crowded fields. Full discussion of issues encountered and therefore potential issues for AO data with large fields of view are found in Section \ref{sec:gc_airopa}, and analysis of irrevocable anisoplanatism is found in Section \ref{sec:aniso}. 

\section{Simulation Methods}
\label{sec:sim_methods}
In order to properly assess the photon noise distribution for each simulated source, we use a Monte Carlo simulation method of measurement over many simulation seeds in order to measure the variation in results, and therefore assess the associated error with a given simulated observation.  In order to obtain relative photometric precision measurements, we apply aperture and PSF-fitting photometry to simulated frames across the varying simulation seeds and calculate values for photometric precision for a variety of stellar magnitudes, field positions, and instrumental modes. Absolute photometric accuracy may then be calculated by comparing the photometric results across simulation seed results to the known simulated flux values adjusted by the instrumental values applicable to each simulation case.

We have catered our simulations to the available PSFs in order to provide a meaningful comparison between PSF-fitting and aperture photometric results.  In Section \ref{sec:fields} we discuss the field configurations used in the simulations for characterizing the photometry error budget for the IRIS imager. For each configuration we limited the field to one imager detector, a 4096x4096 spaxel image at a plate scale of 4 mas per spaxel. For the purposes of exploring the photometric error of IRIS in terms of its absolute accuracy and relative precision, we define the errors associated with a single photometric source using the equations 

\begin{equation}
    \mathrm{Accuracy\ Error\ } (\%) = 100* \frac{\Bar{F} - F_{*}}{F_{*}}
    \label{eq1}
\end{equation}

and
\begin{equation}
    \mathrm{Relative\ Precision\ } (\%) = 100* \frac{\sigma_{Flux}}{\Bar{F}}
    \label{eq2}
\end{equation}

where $\overline{F}$ is the average flux determined through PSF-fitting or aperture photometry, $F_{*}$ is the known seeded flux, and $\sigma_{Flux}$ is the standard deviation of the flux results over all simulation seeds.  Unless otherwise specified, “photometric error” refers to a value of relative precision.

The details of the computer used for these simulations are relevant to the discussion of the photometric results because processing time and CPU draw are a concern for contemporary AO data processing techniques as discussed in Section \ref{sec:gc_airopa}.  We used a 22-core 44-thread processing unit clocked at 2.1GHz and 30 MB Cache, and 250Gb RAM. All seeds for a given simulation were multi-threaded over 25 individual processes at a time. Individual simulation sets were called via scripts as batch files and saved according to their simulation seed number, completely isolating individual seed results in order to provide an accurate sample of the photometric value distribution. 

Each field configuration used for simulations is selected in order to address the photometric effects of specific science questions or instrumental characteristics. Through selecting these configurations we aim to ascertain the photometric performance of IRIS and populate a photometric error budget. When simulating an individual star, we utilize the IRIS data simulator to convolve the estimated PSF (provided by the NFIRAOS team) with flux values estimated for individual filters to simulate the flux distribution of point sources within the imager field of view.  We then estimate instrumental noise properties for each simulated spatial pixel and provide a sky background value based on the simulated filter. The resulting simulated image can then be processed for photometric and astrometric values. 

For each simulated source we conduct aperture and PSF-fitting photometry. Aperture photometry is conducted with the IDL routine Aper, which is a procedure adapted from DAOPHOT and computes concentric aperture photometry at specified field locations using provided aperture radii and conducts sky subtraction. We also use AIROPA to provide photometric and astrometric results via PSF-fitting as discussed in \ref{sec:psf_fitting}.  We provide the locations of each of the seeded sources to both methods of photometry, and calculate independent values of precision for each method of photometric measurement. By comparing the precision results of aperture photometry and AIROPA we gain insight into how AIROPA performs as a function of wavelength, instrumental mode, and scientific field parameters.  This exploration of AIROPA’s performance is a necessary requisite of characterizing IRIS photometric precision and accuracy through PSF-fitting, but not a defined goal of these simulations. For the purposes of ascertaining whether IRIS meets the instrumental photometric requirements, we take the result with better performance in relative precision.

Photometric precision and accuracy is variable based on instrumental modes and field parameters. In order to understand how photometric precision varies with science cases, we define four distinct simulation categories for exploration with the IRIS imager: single-source (single star at the center of the field), grid-source (one star per PSF), binary-case, and crowded field photometry. For each category we run simulations at varying bandpasses and alter simulation characteristics to explore how resulting precision values change with science-field parameters.

\subsection{Observational Parameters}
\label{sec:obs_param}
In order to have consistent comparative values for the photometric performance between simulation cases, we maintain a consistent SNR for each simulation case by varying the total integration time for each simulation. Photometric performance is directly related to  SNR, so in order to explore the effects of varying field configurations and noise properties SNR must be consistent between comparisons. We chose a SNR value of 100 for each of the simulations, and select the integration times shown in Table \ref{tab:itime} for each bandpass in order to maintain this SNR over varying filters.  The integration times shown in Table \ref{tab:itime} are also used in the binary case, with reference to the primary source.

\begin{table}[h]
\begin{center}
\begin{tabular}{c|c|c|c|c|c|c}
     \multirow{2}{*}[-1em]{SNR} & \multirow{2}{*}[-1em]{\tabcell{Magnitude \\ (Vega)}} & \multicolumn{5}{c}{Bandpass Integration Time (Total Seconds)} \\
     & & \tabcell{Z\\$\lambda_{c}$: 0.87 $\mu m$} & \tabcell{Y\\$\lambda_{c}$: 1.01 $\mu m$} & \tabcell{J\\$\lambda_{c}$: 1.24 $\mu m$} & \tabcell{H\\$\lambda_{c}$: 1.62 $\mu m$} & \tabcell{K\\$\lambda_{c}$ 2.19 $\mu m$} \\
    \hline
\multirow{5}{*}{100} & 20 & 36.2 & 32.2 & 26.0 & 38.6 & 85.8 \\
 \cline{2-7}
 & 21 & 91.5 & 82.2 & 67.5 & 120.9 & 263.7 \\
 \cline{2-7}
 & 22 & 234.2 & 214.2 & 182.8 & 454.3 & 965.9 \\
  \cline{2-7}
 & 23 & 615.7 & 587.6 & 543.6 & 2092.1 & 4340.9 \\
  \cline{2-7}
 & 24 & 1719.4 & 1787.8 & 1897.7 & 11254.6 & 22985.4

\end{tabular}

 \caption{Total integration times for the imager by magnitudes and bandpass, zenith value of 30 degrees and 75\% atmospheric turbulence conditions (below average quality seeing). Note: we originally conducted these simulations at SNR of 10 as well, and failed in detection of faint sources using AIROPA, demanding a PSF-fit correlation value of 0.8 and a threshold-above-noise value of 3$\sigma$. \label{tab:itime}}
\end{center}
\end{table}

\subsection{Simulation Seed Analysis}
\begin{figure}[H]
    \centering
    \includegraphics[scale=0.67]{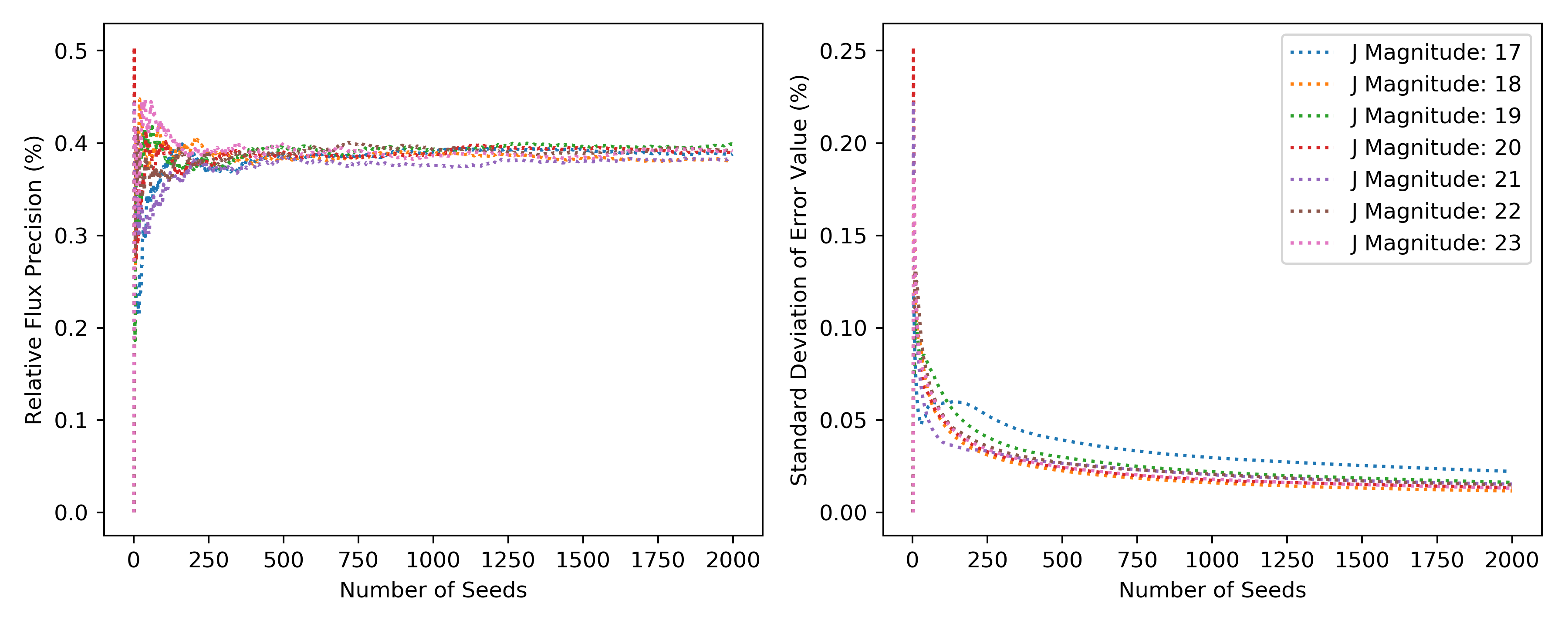}
    \caption{(Left) The relative precision results of a single star simulated in the center of the IRIS field of view as flux is calculated and changes over the simulation seed number.  (Right) The standard deviation of the error values (as shown to the left) as it changes over seed number. These results illustrate the necessity for high numbers of simulations seeds in order to properly assess the error distribution for individual simulations.\label{fig:seed_res}}
\end{figure}

Calculation of the relative photometric precision of the imager requires sampling the distribution of photometric results as those values change due to intrinsic noise for a given simulated case. We achieve this by conducting each simulation case over many individual seeds, and analyzing the resulting values for their error in accordance with equations \ref{eq1} and \ref{eq2}.  In order to ensure that the error distribution is properly sampled by simulating over a sufficient number of simulations seeds, we compare the results of the simulations with how those results change over the seeds accounted for.  This comparison is shown in Figure \ref{fig:seed_res}.

Figure \ref{fig:seed_res} depicts the confidence with which we characterize the error distribution as a function of the number of seeds used in our Monte Carlo simulations. In order to be certain in the error values we report, we spawn 2000 seeds for the purposes of assessing the relative photometric precision for each of the simulation cases described below. 

\section{Simulated Fields \& Results}
\label{sec:fields}

For the purposes of populating the photometric error budget for IRIS/TMT, it is in our interest to understand the contribution of various noise sources to the overall photometric error. This section will define each of the field configurations and simulation parameters adopted in different cases used for assessing the photometric impact of several instrumental and observational parameters.  Because the method of photometric measurement impacts the level of precision and accuracy possible, this section will also discuss the method of photometric measurement. Each of the field configurations below is utilized to investigate the correlation of a noise source to its resulting effect on the photometric error. For each case, a cursory summary of the pertinent results is provided. Because the instrument requirements for photometry are defined in J-band (1.24 $\mu m$), error budget simulations will be reported in J.

\subsection{Single-Source Photometry}
\label{sec:single}
We investigate the photometric performance of a single point source of a certain magnitude located exactly at the center of one imager detector.  For the purpose of comparing photometric performance across wavelength range, we conducted these simulations in the 5 main broadband filters Z (0.87 $\mu m$), Y (1.01 $\mu m$), J (1.24 $\mu m$), H (1.62 $\mu m$), and K (2.19 $\mu m$). We simulated magnitude values from 20 to 24 and calculated integration time to maintain a Signal-to-Noise Ratio (SNR) of 100.  Integration times per magnitude per bandpass are shown in Table \ref{tab:itime}. 

\begin{table}[h]
\begin{center}
\begin{tabular}{c|c|c|c|c|c|c}
     \tabcell{Magnitude \\ (Vega)} & Filter & \tabcell{AIROPA \\ Error (\%)} & \tabcell{Aperture \\ Error (\%)} & Strehl & \tabcell{PSF FWHM \\ (pixels)} & \tabcell{Aperture Radius \\ (Pixels)} \\
     \hline
     \multirow{5}{*}{20} & Z & 0.47 & 0.31 & 0.04 & 2.26 & 8.46 \\
     \cline{2-7}
     & Y &  0.47 & 0.34 & 0.08 & 2.26 & 8.46 \\
     \cline{2-7}
     & J & 0.39 & 0.36 & 0.17 & 2.99 & 9.59 \\
     \cline{2-7}
     & H & 0.35 & 0.61 & 0.32 & 3.91 & 10.72 \\
     \cline{2-7}
     & K & 0.25 & 0.4 & 0.47 & 4.22 & 10.72 \\
     \cline{1-7}
     \multirow{5}{*}{24} & Z & 0.5 & 0.38  & 0.04 & 2.26 & 8.46 \\
     \cline{2-7}
     & Y & 0.51 & 0.82  & 0.08 & 2.26 & 8.46 \\
     \cline{2-7}
     & J & 0.43 & 0.95  & 0.17 & 2.76 & 9.59 \\
     \cline{2-7}
     & H & 0.4 & 1 & 0.32 & 3.91 & 10.72 \\
     \cline{2-7}
     & K & 0.31 & 0.54  & 0.47 & 4.22 & 10.72
\end{tabular}
 \caption{The simulation results for a single point source at the center of one imager detector field of view. The location for the simulated source in reference to the coordinates defined in Figure \ref{fig:iris_layout} is (8.8, 8.8) in arcseconds off-axis. As all of the simulations in this paper, these were conducted using a zenith angle of 30$^\circ$ and "poor" atmospheric conditions in order to provide a conservative estimate for the photometric error.  \label{tab:single_res}}
\end{center}
\end{table}

The photometric error results from the single-source case are shown in Table \ref{tab:single_res}.  These simulations correspond to a single star simulated at the (8.8", 8.8") point as defined in Figure \ref{fig:iris_layout}. These values serve as a consistent base error value for comparison with more complicated field configurations. Additionally, we use this same configuration to test the effects of different instrumental noise values on photometric error by altering the instrumental values for read noise and dark current in the data simulator.

\subsubsection{Instrumental Noise Characteristics}
We test the contribution of read noise and dark current to photometric error by conducting two sets of simulations for each with altered values in addition to our standard simulation, in which we use the expected values of read noise and dark current achieved by Teledyne Hawaii-2RG and Hawaii-4RG detecotrs. We simulate values for each at twice the standard values, and also lacking those noise attributes entirely.  The results of these simulations are shown in Table \ref{tab:inst_noise} along with the standard deviations of the error values over all simulation seeds.

\begin{table}[h]
\begin{center}
\begin{tabular}{c|c|c|c}
\tabcell{Magnitude \\ (J)} & Noise Condition & \tabcell{AIROPA \\ Error (\%)} & \tabcell{Standard Deviation \\ of Error over all Seeds} \\
\hline
\multirow{5}{*}{20} & Standard & 0.39 & 0.01 \\
\cline{2-4}
& Read Noise: 0 & 0.39 & 0.03 \\
\cline{2-4}
& Read Noise: 10 $e^{-}$ & 0.41 & 0.01 \\
\cline{2-4}
& Dark Current: 0 & 0.39 & 0.01 \\
\cline{2-4}
& Dark Current: 0.004 $e^{-}/s$ & 0.39 & 0.01 \\
\hline
\multirow{5}{*}{21} & Standard & 0.38 & 0.01 \\
\cline{2-4}
&Read Noise: 0 & 0.38 & 0.02 \\
\cline{2-4}
&Read Noise: 10 $e^{-}$ & 0.40 & 0.02 \\
\cline{2-4}
&Dark Current: 0 & 0.38 & 0.02 \\
\cline{2-4}
&Dark Current: 0.004 $e^{-}/s$ & 0.41 & 0.02 \\
\hline
\multirow{5}{*}{22} & Standard & 0.39 & 0.01 \\
\cline{2-4}
& Read Noise: 0 & 0.40 & 0.01 \\
\cline{2-4}
& Read Noise: 10 $e^{-}$ & 0.39 & 0.02 \\
\cline{2-4}
& Dark Current: 0 & 0.41 & 0.01 \\
\cline{2-4}
& Dark Current: 0.004 $e^{-}/s$ & 0.40 & 0.02
\end{tabular}
\caption{Differing noise profiles used for characterizing instrumental noise characteristics.  The standard noise values consist of 5 $e^{-}$ of read noise and 0.002 $e^{-}/s$ for dark current.\label{tab:inst_noise}}
\end{center}
\end{table}

The results in Table \ref{tab:inst_noise} show that the instrumental noise contributions of read noise and dark current insofar as they pertain to the effects on photometric error in J-band are small compared to the total noise fluctuations.  The differences between each set of the photometric error results, compared with the standard deviation of the relative error values over 2000 seeds, are illustrative of the fact that the contributions of dark current and read noise to photometric error are very small. For the purposes of the error budget, these values can be assessed as $<$0.01\% inferred from the noise distribution of these results.

\subsection{Grid-Source Photometry}
\label{sec:grid_source}
In order to estimate the effect that PSF-variability has on photometric precision (absent of the effects of overlapping PSFs), we simulate a grid of point sources in the field at the locations for which we have a corresponding PSF.  These locations are shown in Figure \ref{fig:iris_layout}, relative to an on-axis location of (0”, 0”).  We simulated a slightly larger field of view (in order to encompass the PSFs corresponding to the edges of the field of view), and simulated a point source at each of the locations of the 5x5 grid in each of the 5 main filters, at the same magnitudes and integration times as shown in Table \ref{tab:itime}.

\begin{figure}[H]
    \centering
    \includegraphics[scale=0.9]{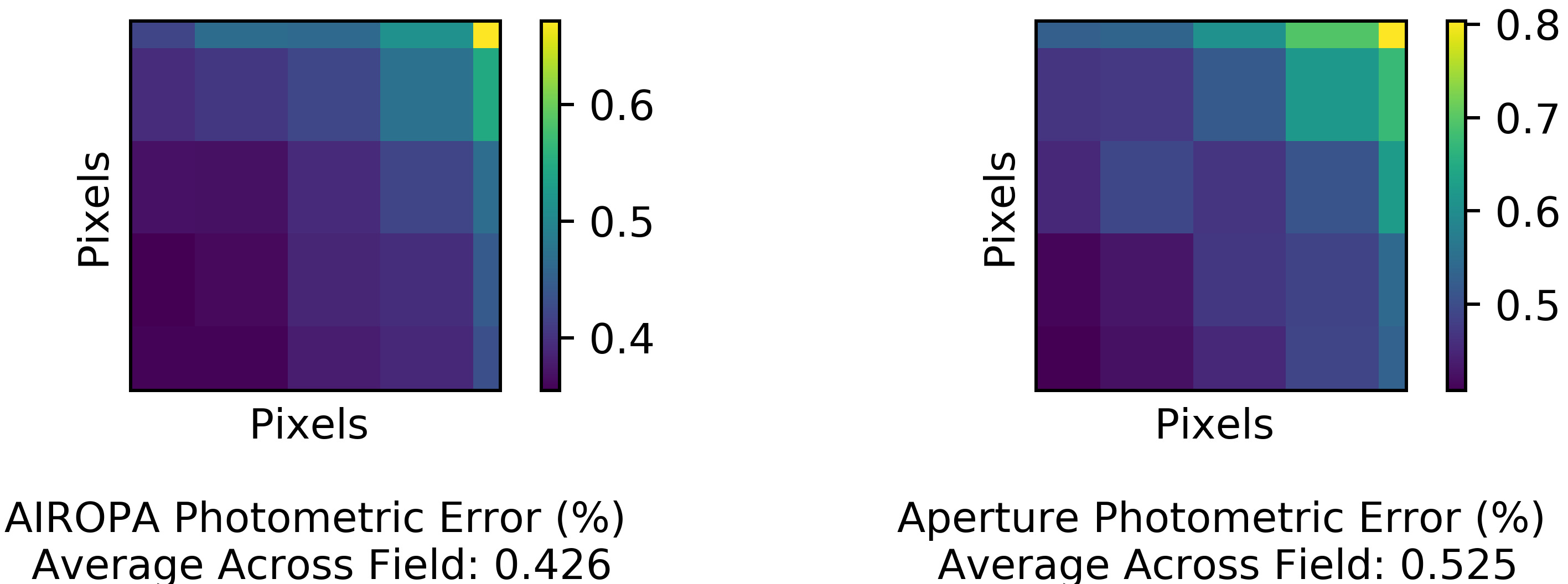}
    \caption{(Left) Relative photometric precision results via AIROPA as it varies across the field simulated with the 5x5 grid of PSFs shown in Figure \ref{fig:iris_layout}. (Right) Relative photometric precision using aperture photometry over the same sources, with aperture radii equal those defined in Table \ref{tab:single_res}.  These figures represent the results of the J-band grid-source simulation with point sources of 22nd magnitude. \label{fig:grid_res}}
    
\end{figure}

This grid of simulated point sources allowed us to investigate the effect of variation of the PSF on photometry, utilizing AIROPA. The simulations at these points result in an error value corresponding to each field location.  This yields an error ``heat map", which illustrates the effect of variable PSFs on photometric error results. The J-band results of this grid simulation are shown in Figure \ref{fig:grid_res}. The spatial variance to the error is consistent with the change in Strehl ratio across the field of view shown in Figure \ref{fig:iris_layout}.

\subsection{Binary-Source Photometry}
\begin{figure}[H]
    \centering
    \includegraphics[scale=0.70]{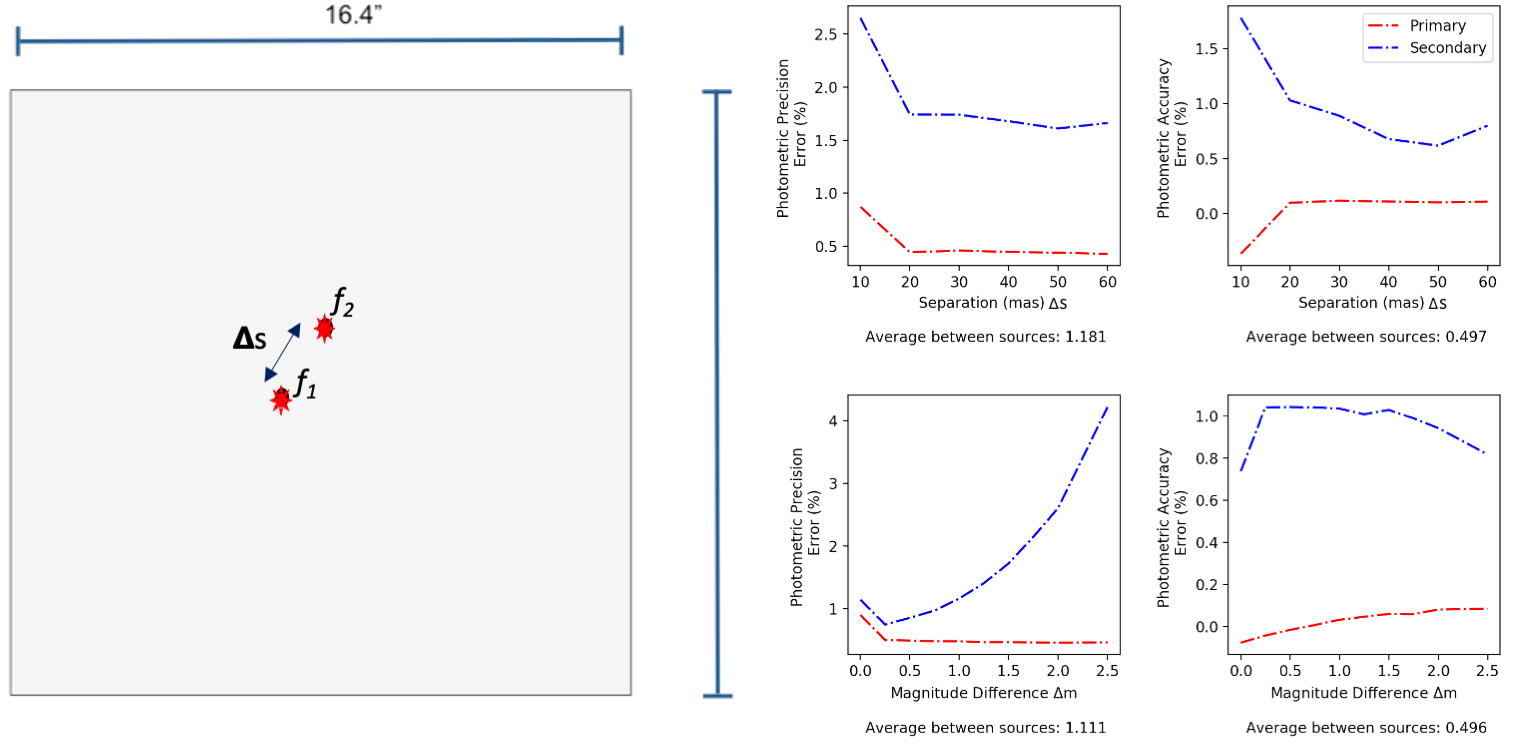}
    \caption{(Left) The field configuration of the simulated binary stars. We azimuthally averaged the position of the secondary source in order to account for asymmetrical PSF effects.  (Right) Photometric precision (left column) and accuracy (right column) as it changes over distance between the binary stars (upper row) and as it changes over difference in magnitude between the two sources (lower row).\label{fig:binary_res}}
\end{figure}

For binary-case photometry we simulated two sources in the center of one imager detector field of view, varying the separation and flux difference between them.  We simulated in J, with a primary source magnitude of 20 and varied the secondary source magnitude with $\Delta$m=0.25 steps. We averaged the location of the secondary source azimuthally over the simulation seeds, in order to compensate for asymmetrical PSF effects on the returned flux.  The configuration used and results for the two sources in the binary case are shown in Figure \ref{fig:binary_res}.

The binary simulation results emphasize the importance of distinguishing sources in achieving high photometric precision; when two stars are proximal on the sky, their PSFs blend and may become indistinguishable using PSF-fitting. We will refer to this effect as source confusion. The relative precision achieved as it varies over star separation in right upper-left Figure \ref{fig:binary_res} shows that a separation of 20 mas is necessary to avoid a decrease in photometric precision. These simulations also demonstrate that the effect of overlapping halos contributes to photometric error as much as $\sim$1\%.

\subsubsection{PSF-R}
\label{sec:PSFR}
The binary system case is useful for assessing the effect of overlapping PSFs on photometric error.  In this case, knowledge of the PSF becomes more important than in the isolated star field configuration, making it a good candidate for testing the veracity of the reconstructed PSFs. The comparative results of the same binary source simulation conducted using the package of reconstructed PSFs and science PSFs is shown in Table \ref{tab:res_PSF-R}.  

\begin{table}[h]
\begin{center}
\begin{tabular}{c|c|c|c|c|c|c|c|c|c|c|c|c}
    \multirow{3}{*}{$\Delta$m (K)} & \multicolumn{4}{c|}{\tabcell{Separation:\\ 0.01"}} & \multicolumn{4}{c|}{\tabcell{Separation:\\ 0.02"}} & \multicolumn{4}{c}{\tabcell{Separation:\\ 0.03"}} \\
    \cline{2-13}
     & \multicolumn{2}{c|}{\tabcell{AIROPA\\ Error (\%)}} & \multicolumn{2}{c|}{\tabcell{Aperture \\ Error (\%})} & \multicolumn{2}{c|}{\tabcell{AIROPA\\ Error (\%)}} & \multicolumn{2}{c|}{\tabcell{Aperture \\ Error (\%})}& \multicolumn{2}{c|}{\tabcell{AIROPA\\ Error (\%)}} & \multicolumn{2}{c}{\tabcell{Aperture \\ Error (\%})} \\
    \cline{1-13}
    & SCI & PSF-R & SCI & PSF-R & SCI & PSF-R & SCI & PSF-R & SCI & PSF-R & SCI & PSF-R \\
    \cline{1-13}
    0.0 & 24.2 & 25.5 & 1.14 & 0.95 & 1.18 & 1.39 & 1.40 & 1.54 & 1.65 & 1.77 & 2.14 & 2.23 \\
    \cline{1-13}
    0.25 & 27.5 & 27.6 & 1.50 & 1.69 & 1.43 & 1.56 & 1.55 & 1.71 & 1.81 & 2.04 & 2.14 & 2.33 \\
    \cline{1-13}
    0.5 & 29.7 & 29.8 & 1.17 & 0.96 & 1.25 & 1.43 & 1.36 & 1.52 & 1.83 & 1.92 & 2.07 & 2.14 \\
    \cline{1-13}
    0.75 & 29.2 & 29.7 & 1.51 & 1.68 & 1.52 & 1.65 & 1.60 & 1.75 & 2.11 & 2.35 & 2.16 & 2.36 \\
    \cline{1-13}
    1.0 & 24.4 & 21.8 & 1.16 & 0.98 & 1.40 & 1.65 & 1.38 & 1.55 & 2.27 & 2.38 & 2.10 & 2.18 \\
    \cline{1-13}
    1.25 & 29.0 & 26.4 & 1.53 & 1.72 & 1.79 & 1.94 & 1.61 & 1.76 & 2.73 & 2.95 & 2.21 & 2.42 \\
    \cline{1-13}
    1.5 & 38.5 & 38.5 & 1.18 & 0.99 & 1.81 & 2.08 & 1.40 & 1.57 & 3.09 & 3.23 & 2.17 & 2.23 \\
    \cline{1-13}
    1.75 & 49.0 & 49.3 & 1.53 & 1.72 & 2.27 & 2.43 & 1.63 & 1.81 & 3.84 & 4.06 & 2.28 & 2.47 \\
    \cline{1-13}
    2.0 & 49.4 & 50.7 & 1.19 & 1.00 & 2.58 & 2.86 & 1.42 & 1.61 & 4.47 & 4.67 & 2.22 & 2.29 \\
    \cline{1-13}
    2.25 & 41.1 & 44.3 & 1.50 & 1.71 & 3.32 & 3.49 & 1.66 & 1.81 & 5.59 & 5.85 & 2.32 & 2.54 \\
    \cline{1-13}
    2.5 & 31.2 & 33.7 & 1.17 & 1.00 & 3.50 & 3.79 & 1.44 & 1.62 & 6.61 & 6.91 & 2.28 & 2.36 
\end{tabular}
 \caption{The PSF-R comparison simulation results for the binary case for different separation values between sources and magnitude differences from the primary source. Here "SCI" refers to simulations conducted using the same "Science PSF" for extraction as was used to simulate each star, and "PSF-R" refers to those in which an estimated reconstructed PSF was used for extraction. These simulations were conducted using the same atmospheric conditions and zenith angle as those defined in \ref{sec:single}. Due to restrictions with the PSFs available in our datasets, we simulated these only in K-band. \label{tab:res_PSF-R}}
\end{center}
\end{table}

The error values at a separation of 10 mas are dominated by source confusion, which is more exaggerated for the PSFs used in the R-PSF comparison simulations than those illustrated in Figure \ref{fig:binary_res}. This is illustrated clearly through the comparison with the aperture results, which are lower in error because the aperture for the 10 mas case is large enough to contain the cores of both simulated stars. Because we wish to assess the difference in performance of our PSFs, we exempt the 10 mas case and consider only the non-confused results. From these selections (separations of 20 - 60 mas), we then assess the associated error of PSF-Reconstruction by taking the difference between our separate PSF error results in quadrature, and averaging them over the magnitude differences.  From this we assess the impact of PSF-reconstruction on photometric error to be of the order of $\sim$1\%.

\subsection{Crowded Field Photometry}
\label{sec:crowded_field}
The current photometry requirements for the IRIS instrument are to be applied in the case of a moderately crowded field, as defined here. This section will discuss the purpose of using the crowded field as the basis for defining whether the IRIS imager meets its instrument requirements, and describe the specifics of the field configurations used in the photometry simulations.

The purpose of conducting crowded-field simulations is to measure the resulting change in photometric error due to the effects of overlapping halos of point spread functions and crowded sources. However, it is difficult to separate the instrumental effects from the resulting difficulty in photometric method, because close sources and overlapping PSFs correspondingly create difficulty in PSF-fitting routines. Therefore, the aspects of IRIS we hope to test the effects on photometric error by simulating the crowded-field configuration are PSF-variability across the field (anisoplanatism), effects of overlapping PSF halos, a large range of magnitudes of sources (contrast ratio), and the capabilities of current tools to recover sources.

For the photometry error budget we define the case for a crowded field which is not largely impacted by source confusion. The extreme case in which all of the listed field characteristics are more exacerbated (many sources, confusion, etc) is discussed in section \ref{sec:gc_airopa} through simulation of the Galactic Center. The field configuration used for the crowded field is the described 16.4" x 16.4" (4096 x 4096 pixels) field of view, in J-band, at a zenith angle of 30 degrees, and atmospheric quality of best 75\%.  We simulate 1000 stars within the field randomly distributed, with a minimum distance between stars equal to three times the full-width half-maximum of the PSF ($\sim$8.1 pixels or 32.4 mas, for the 2.7 pixel or 10.8 mas full-width half-maximum of the J-band PSF at 4 mas per spatial pixel). The luminosity of the stars within the field is a uniform distribution of source magnitudes between 20 and 24, with an integration time which achieves a SNR of 100 for the mean source brightness in the field ($\sim$196 seconds for J-band). We assume only a single calibration star within the field of view, which translates to having a minimum of 0\% error in the error heat map for the simulated field. 

\begin{figure}[h]
    \centering
    \includegraphics[]{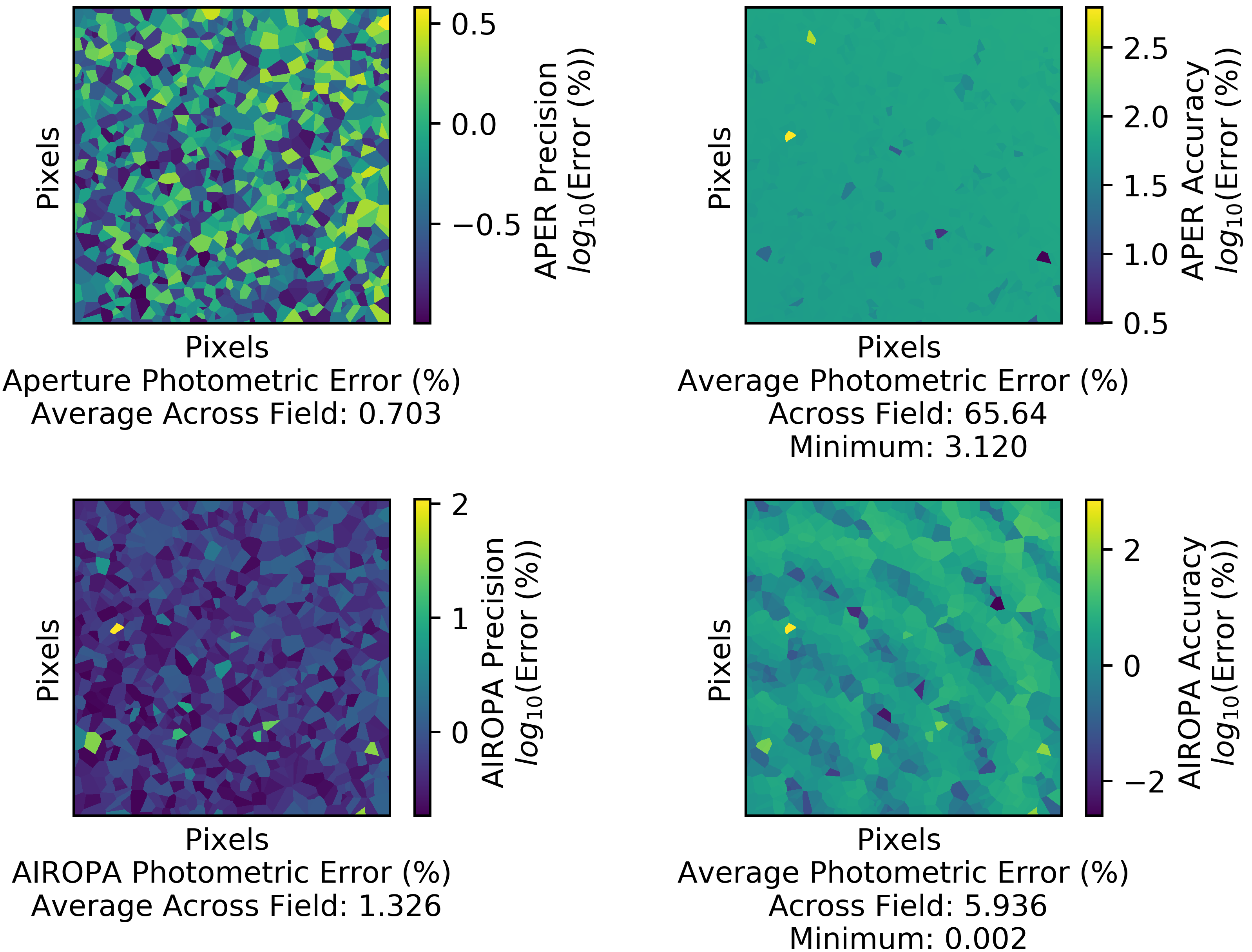}
    \caption{(Upper Row) We show the photometric precision (left) and accuracy (right) over the field as returned from aperture photometry.  (Lower Row) We show the photometric precision (left) and accuracy (right) achieved by AIROPA. Figures are log-scale to compensate for outliers and to better illustrate the spatial structure of the AIROPA accuracy (lower right).\label{fig:crowded_res_5}}
\end{figure}

As in Section \ref{sec:grid_source}, the results of the crowded field simulation are interpreted via an error heat map, in which there is an associated photometric precision and accuracy term for each star in the field. Using the 5x5 grid of PSFs shown in Figure \ref{fig:iris_layout} as the field-variable input PSF-grid for our AIROPA PSF-fit, we achieve the results shown in Figure \ref{fig:crowded_res_5}.

\begin{figure}[h]
    \centering
    \includegraphics[]{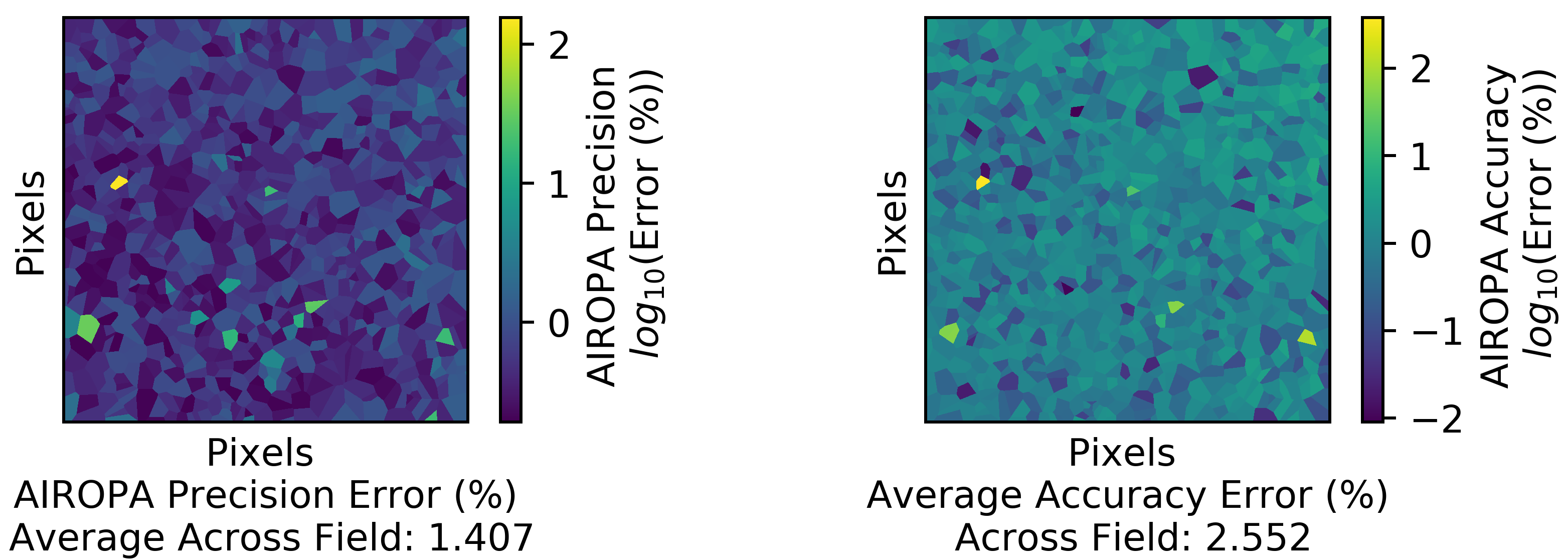}
    \caption{We show the photometric precision (left) and accuracy (right) achieved by AIROPA. Figures are log-scale to compensate for outliers and to better illustrate the spatial structure of the AIROPA accuracy. This plot was generated in the same manner as Figure \ref{fig:crowded_res_5}, but used an interpolated 15x15 PSF grid for fitting across the field in AIROPA, resulting in improved consistency in the photometric accuracy across the field as compared with the 5x5 PSF grid defined in Figure \ref{fig:iris_layout}. \label{fig:crowded_res_15}}
\end{figure}

Close inspection of the AIROPA accuracy results in Figure \ref{fig:crowded_res_5} yields repeating spatial structure to the error which is associated with the input PSF-grid used for fitting via AIROPA. This repeating error structure is due to the differences between the interpolated PSF used for the simulation of each source and the "nearest neighbor" approach used by AIROPA and the input 5x5 PSF grid defined in Figure \ref{fig:iris_layout}.  Because the error budget assumes only a single calibration star within the field, it is of interest to minimize the spatial dependencies of the error.  This is achieved by interpolating the input PSF grid to more finely sample that PSFs across the field.  By interpolating a PSF to a 15x15 grid across the imager field of view, we yield improved results as shown in Figure \ref{fig:crowded_res_15}.

As compared with the AIROPA accuracy results of Figure \ref{fig:crowded_res_5}, Figure \ref{fig:crowded_res_15} demonstrates an improvement across the field of the spatial consistency of the photometric error.  This therefore improves the achievable photometric error to be assessed for the photometric error budget, because a single calibration star within the field cannot account for spatially variable inaccuracies due to spatially dependent PSFs.  We found finer spatial sampling of the PSF grid yielded minimal improvements. These differences exemplify the importance of properly sampling the PSF across the field in accordance with the severity of the PSF variability. 

We assume a single calibration star within the imager field of view to populate the photometric budget. While uniform deviations in photometric accuracy can be compensated for by appropriate calibration stars, this does not necessarily account for the spatial variations in accuracy shown in Figures \ref{fig:crowded_res_5} and \ref{fig:crowded_res_15}. Since the minimum in the accuracy error heat map is near zero and there is little uniformity in the error across the field, it is unlikely that further calibration would necessarily improve photometric error. Therefore, in order to assess a conservative estimate for the photometric error achieved, we adopt the value estimated by the photometric accuracy simulations and asses the error to be $\sim$2.6\%.

\subsubsection{Confusion}
\label{sec:confusion}

\begin{figure}[h]
    \centering
    \includegraphics[scale=0.45]{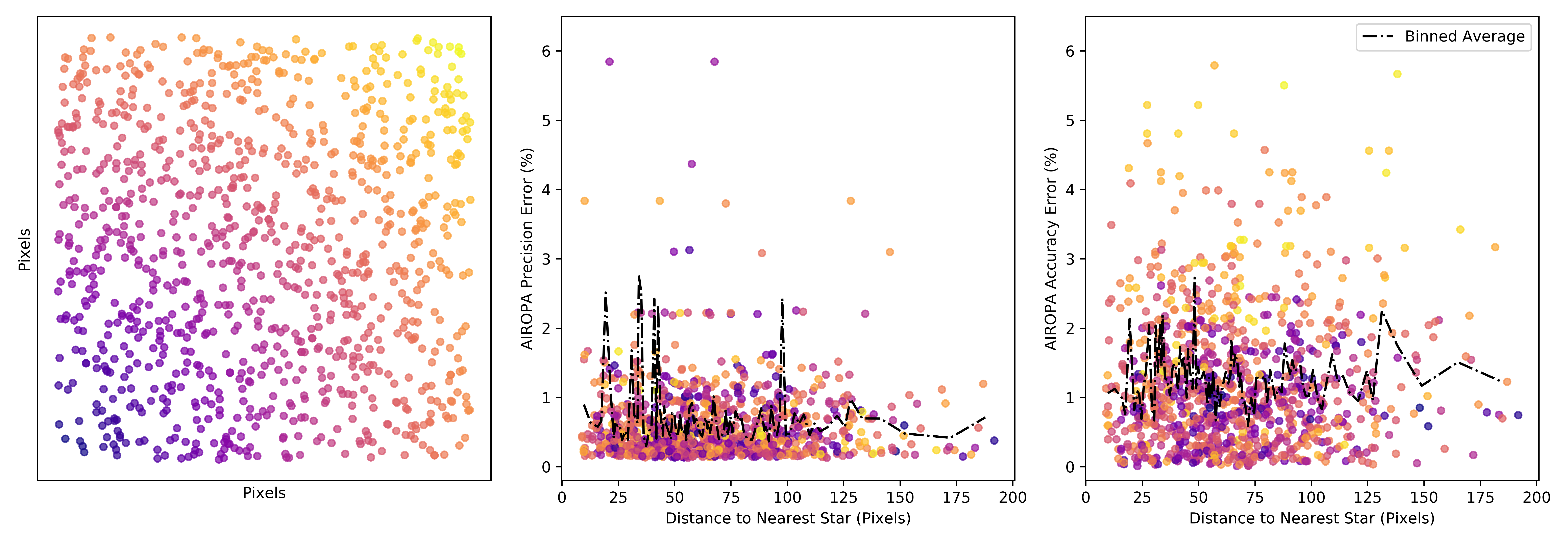}
    \caption{(Left) the simulated crowded field colored in accordance to distance from the on-axis location (0.0",0.0"), which corresponds to lower Strehl ratio as illustrated in Figure \ref{fig:iris_layout}.  (Middle) Each star's photometric precision error as it corresponds to the distance to its nearest star neighbor. (Right) Each star's photometric accuracy error as it corresponds to the distance to its nearest star neighbor. The binned average (using a bin size of 10 pixels) is shown to highlight the lack of a trend upward as distance to the star's nearest neighbor approaches 0, which would indicate the presence of source confusion in our photometric measurements. \label{fig:confusion}}
\end{figure}

We define source confusion as the blending of proximal source PSFs such that distinguishing the stars in PSF-fitting becomes problematic. The effects of confusion on photometric error are non-negligible and for the purposes of the photometry budget are not considered. We verify that confusion does not contribute to the assessed photometric error from the crowded field simulation by estimating the error as a function of each star's distance from its closest neighboring star. The scatter plot of simulated sources with color correlating to distance from the on-axis point is displayed in Figure \ref{fig:confusion}.

\subsubsection{Anisoplanatic Effect}
\label{sec:aniso}

\begin{figure}[h]
    \centering
    \includegraphics[scale=0.45]{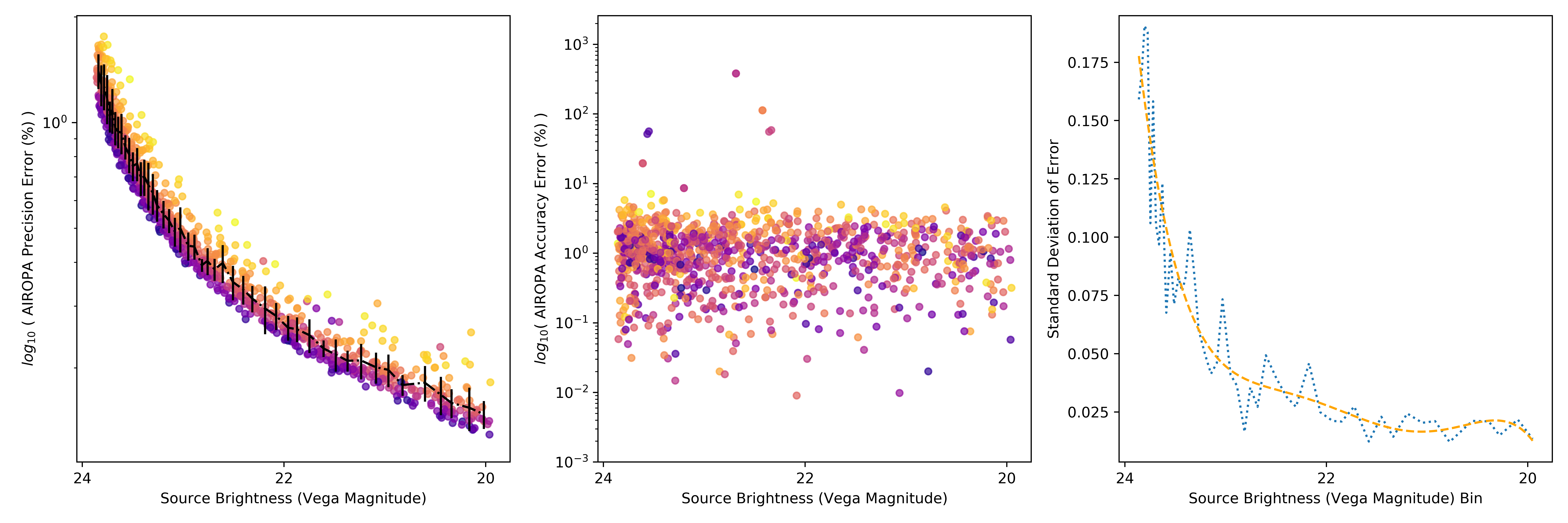}
    \caption{Demonstration of the effect of anisoplanatism on photometric error. (Left) The photometric precision for each source plotted and colored according to the scatter plot of Figure \ref{fig:confusion} distributed over magnitude. Plotted on a log scale, the effect of anisoplanatism is clear due to the separation of sources vertically in accordance with their distance from the on-axis point (0.0", 0.0").  Bars defining the standard deviation for each magnitude bin are included to highlight the values of the rightmost plot. (Middle) The photometric accuracy for each crowded field source distributed over magnitude.  As opposed to the precision value, the effect of anisoplanatism is less pronounced, and the effect of anisoplanatism on photometric accuracy is better demonstrated by Figure \ref{fig:confusion}. (Right) The plotted standard deviation of each magnitude bin in (Left) as marked by the black bars. \label{fig:aniso}}
\end{figure}

Anisoplanatism refers to changing atmospheric turbulence across the field of view; even assuming perfect performance of NFIRAOS, the corrections of the AO system will desynchronize from atmospheric effects as distance from the corrected field location increases, which results in variation of the PSF across the field. The effects of anisoplanatism on photometry is of particular interest because it represents the performance of AIROPA in compensating for the spatially dependent PSF. Using the color distribution of the simulated star scatter plot in Figure \ref{fig:confusion}, we estimate the contribution of anisoplanatism to photometric error by analyzing the photometric error distribution as a function of star magnitudes, as shown in Figure \ref{fig:aniso}. From the variance of photometric error values within individual brightness bins illustrated in Figure \ref{fig:aniso} we assess the effect of anisoplanatism for high SNR sources to be $\sim$0.025\%.

\subsection{Galactic Center}
\label{sec:gc_airopa}
To simulate an extreme crowded field case we use the Galactic Center (GC) field as our test-bed since it is a scientific case which will fully take advantage of the capabilities of IRIS. The GC field configuration allows us to further explore our current photometric methods directly with a science case. The results from the crowded-field configuration are not analogous to the different field simulations conducted thus far, because it is difficult in the Galactic Center case to find direct precision values for each source in the field. This is true particularly because the sources recovered are heavily dependent on the PSF-fitting parameters of AIROPA, and require a high number of CPU-processing hours to simulate.

The GC is simulated with a field of view of 16.4\arcsec x16.4\arcsec (4096x4096 pixels) centered on Sgr A*, K-band imaging filter, zenith angle of $30^{\circ}$, 2.2 second integration time (the presumed instrument minimum), and with a minimum simulated stellar magnitude of $m_K$=27 (Vega). The luminosity profile of the simulated field is displayed in Figure \ref{fig:gc_recovery}, and underestimates the actual number of sources expected in the GC in order to reduce CPU-processing time in our simulations. The number of stars seeded in the GC simulation includes known sources down to 23rd magnitude, with additional sources down to 27th magnitude randomly dispersed.  The total number of sources simulated is 130200, a large underestimate of the number that will likely be observed in the GC following the advent of observations by 30-meter-class telescopes\cite{Do2019}.

In simulating the Galactic Center we have encountered several issues specific to the processing of data of dense stellar fields with large fields of view and high sensitivities. Because the GC is an extreme case which tests the level to which we are capable of processing and recovering faint and confused sources, the recoverable photometric values are highly dependent on the specific methods with which we process the simulated frames.  The PSF-fitting requisite correlation value for a source (‘C’) and the requisite threshold-above-noise value (‘threshold’) used in PSF-fitting by AIROPA both have a large contribution to the quality and quantity of recovered sources. Additionally, because there are several hundreds of thousands of sources within the field, the processing time is of large concern when studying the GC field.  These issues are discussed in detail below.

\subsubsection{Problems of CPU Hours and Source Recovery}
\begin{figure}[H]
    \centering
    \includegraphics[scale=0.5]{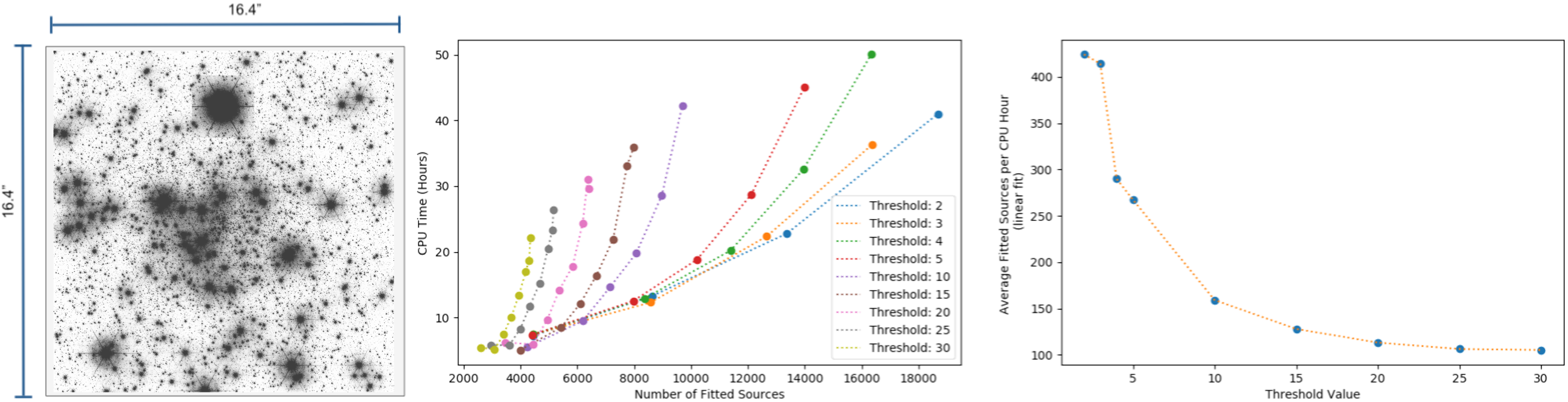}
    \caption{(Left) The simulated Galactic Center field, centered on Sagittarius A*. (Middle) CPU processing time versus source recovery via AIROPA with different threshold values, where each point ascending a given threshold line is a decreasing correlation value for that simulation. (Right) Average total number of sources returned by AIROPA per CPU hour for a given set of simulations with consistent threshold values (and varying correlation values). \label{fig:gc_field}}
\end{figure}

\begin{figure}[H]
    \centering
    \includegraphics[scale=0.5]{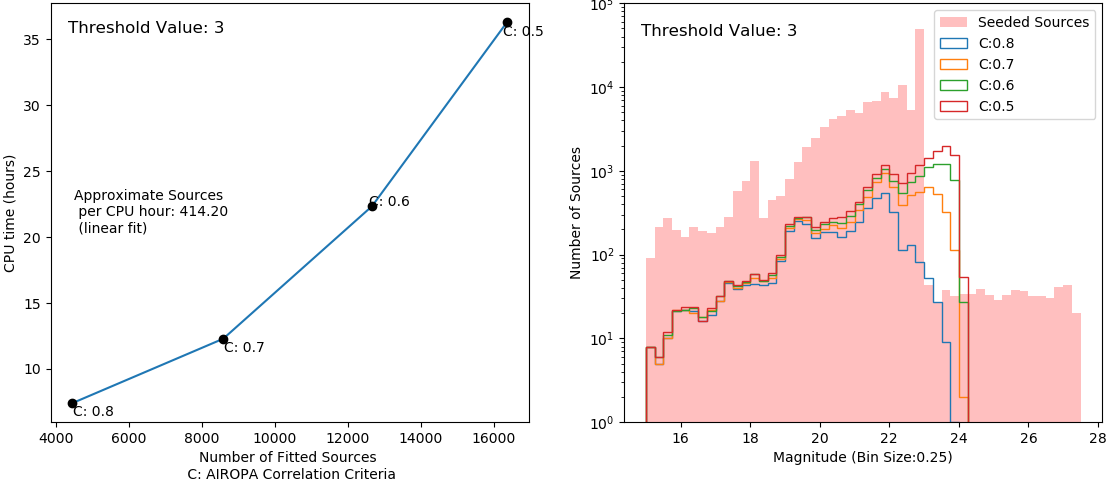}
    \caption{(Left) The computing time over different PSF-correlation requirement values for AIROPA. (Right) the source recovery for each correlation value specified for the given simulation plotted against the luminosity profile of the seeded sources. \label{fig:gc_recovery}}
\end{figure}

The performance of AIROPA in the GC is heavily dependent on the provided requisite correlation value for PSF-fitting - the assigned value for which a given source must match the corresponding PSF in order to be returned as a source. Source recovery is also defined by the provided threshold value - the value above the standard noise which a source must meet in order to be analyzed in the PSF-fitting routine. Both the correlation value and requisite threshold impact the number of sources recovered and the overall CPU time which is required for PSF-fitting. These CPU hours refer to the time required for AIROPA to complete PSF-fitting iterations and return photometric and astrometric values for a single simulation. Figures \ref{fig:gc_field} and \ref{fig:gc_recovery} illustrate the relationship between correlation value, threshold value, CPU time, and source recovery.

The Galactic Center poses a challenge in computing, for appropriately fitting the great number of sources in the field and achieving high source recovery.  Because of the high sensitivity of IRIS and the large aperture of TMT, the number of sources observed in the GC will be even greater than the number simulated here. Figure \ref{fig:gc_field} shows that AIROPA improves in fitting efficiency per CPU hour at lower threshold values, despite the increased total computing time. We select a threshold value of 3 to demonstrate the recovery achieved for our simulated field, and find the results for varied PSF-fit correlation requirements in Figure \ref{fig:gc_recovery}.

The source recovery percentage of the simulation results with correlation values of 0.8, 0.7, 0.6, and 0.5 are then 3.4\%, 6.6\%, 9.7\%, and 12.6\% respectively.  This low recovery rate cannot be improved by decreasing the requisite PSF correlation due to the corresponding negative impacts to accuracy shown in Figure \ref{fig:gc_recovery} (Right). Conducting high-recovery, high-accuracy GC field measurements, additionally with practical processing time requirements, will be a challenge for GC astronomers using 30-meter-class telescopes. As discussed in Section \ref{sec:sim_methods}, the computer used for these simulations was not limited in terms of CPU core count. Higher single-core processing clock speeds may prove marginally beneficial for improving processing time, but likely not to the extent which would enable complete source recovery at high accuracy. We therefore conclude that advancing our processing algorithms is a priority for the development of this science case. Future improvements in PSF-fitting algorithms and faster computations for iterative PSF-fitting will be a requirement for dense stellar field data processing with high-sensitivity and large field of view instruments such as IRIS. 

\section{Summary}

We have reported the results of the simulations conducted to ascertain the photometric performance of the IRIS instrument in conjunction with NFIRAOS/TMT. Using the full IRIS data simulator and large package of simulated NFIRAOS-provided PSFs, we have conducted large Monte Carlo simulations to sample the value distributions of the projected performance of IRIS using aperture and spatially-variable PSF-fitting photometry. We have simulated thousands of seeds for the field configurations of the single-star, binary-star, and crowded field cases. We have shown using these configurations that for the purposes of the instrument photometry requirements, we achieve an accuracy level of $\sim$2.6\% for simulations with a mean SNR of 100 in J-band ($\lambda_{c}$: 1.24 $\mu m$), and assess the contributions to this value of the instrumental noise characteristics, effects of overlapping PSF halos, and anisoplanatism. We have assessed the contributions to photometric error of current PSF-R algorithms in comparison with the best estimate of on-sky PSFs and calculate an estimate of $\sim$1\%. We have also explored the Galactic Center as a prospective science case as it relates to advancements in data image processing and analyzed the current problems to crowded field photometry that such an extreme case poses to astronomers. 

\acknowledgments     
 
The TMT Project gratefully acknowledges the support of the TMT collaborating institutions. They are the California Institute of Technology, the University of California, the National Astronomical Observatory of Japan, the National Astronomical Observatories of China and their consortium partners, the Department of Science and Technology of India and their supported institutes, and the National Research Council of Canada. This work was supported as well by the Gordon and Betty Moore Foundation, the Canada Foundation for Innovation, the Ontario Ministry of Research and Innovation, the Natural Sciences and Engineering Research Council of Canada, the British Columbia Knowledge Development Fund, the Association of Canadian Universities for Research in Astronomy (ACURA), the Association of Universities for Research in Astronomy (AURA), the U.S. National Science Foundation, the National Institutes of Natural Sciences of Japan, and the Department of Atomic Energy of India.


\bibliographystyle{spiebib}
\bibliography{report.bib}   

\end{document}